# Degradation of superconducting properties in MgB$_2$ films by exposure to water

**H Y Zhai, H M Christen, L Zhang, M Paranthaman, P H Fleming and D H Lowndes**

Oak Ridge National Laboratory, Oak Ridge, TN 37931-6056

**Abstract**
The effect of water exposure on MgB$_2$ is studied by submerging an 800 nm thick MgB$_2$ film into deionized water at room temperature for 1 hour, 4 hours, 10 hours, and 15 hours, and by analyzing the resulting material using scanning electron microscopy and resistance vs. temperature measurements. It is clearly observed that $T_c^{onset}$ of these films (obtained by an *ex-situ* reaction of a e-beam evaporated boron layer) remains unchanged throughout this process, indicating that at least a portion of the sample retains its original bulk-like properties. The data is consistent with an interpretation in which a portion of the exposed film – likely the region closest to the substrate – becomes superconducting only at ~ 25 K. It is possible that this low-$T_c$ region already exists in the as-prepared film, and we observe that its $T_c$ coincides with that of MgB$_2$ films obtained by annealing precursor films prepared by pulsed laser deposition. Therefore the data presented here not only illustrate the degradation of MgB$_2$ in water but also shed light onto the differences and similarities between films obtained via different routes.

## 1. Introduction

The discovery of superconductivity at 39K in the simple binary magnesium diboride (MgB$_2$) [1,2], almost doubling the $T_c$ within the family of classic superconductors, has brought great excitement to the condensed-matter community. Today, applications of copper-oxide based high-$T_c$ materials are complicated by technological hurdles related to grain boundary transport and cost-effective production of large quantities of these complex cuprates. In contrast, the grains in MgB$_2$ polycrystalline samples are strongly linked [3,4], and the starting materials are readily available. In addition to the great promise for applications, the fact that this layered material, with conduction predominantly occurring within the graphite-like B-layers, in many ways behaves similarly to a classic BCS superconductor, provides a challenge to our theoretical understanding of both MgB$_2$ and cuprate superconductors [5].

Experimental efforts have focused on wires [6] and thin films [7-14] with promising results obtained within weeks of the original discovery.

Materials stability is an issue of great concern for various superconducting device applications, and most high-temperature superconductors are highly sensitive to water and moist air [15-18]. It is thus very important to explore the influence of exposure of MgB$_2$ to air and water both with respect to applications and in order to gain a more fundamental understanding of its physical and chemical properties. In previous publications[7-11] we have reported the fabrication of MgB$_2$ films by different procedures and compared the growth mechanisms for films formed via an *ex-situ* reaction of e-beam evaporated boron ($T_c$ ~ 39 K) or via *in-situ* annealing of PLD-grown precursor layers ($T_{c0}$ ~ 25).

## 2. Experimental

In this report, we present a study of the influences of exposure to water of MgB$_2$ films prepared via the e-beam evaporation/*ex-situ* annealing route. The results are consistent with the observation of two distinctively separate layers, namely a top layer of well-crystallized MgB$_2$ grains containing very few impurities and exhibiting bulk-like superconducting properties, and a second layer closer to the substrate showing evidence of oxygen contamination [10]. The present study sheds light onto the properties of these distinctively different layers and point out similarities with PLD-grown films.

The MgB$_2$ film used in this study was prepared by e-beam evaporation of a 500 nm thick B film onto a R-plane Al$_2$O$_3$ substrate, followed by a 20-minute anneal at 890°C in a crimped Ta cylinder containing additional MgB$_2$ and Mg and placed inside a sealed quartz tube. The resulting 800 nm thick film was characterized by a $T_{c0}$ > 30 K, with a $T_c^{onset}$ ≈ 39 K. Most of the MgB$_2$ films studied in our group exhibit much sharper transitions with similar $T_c^{onset}$. The choice of this particular sample for the water-exposure study was fortunate as it provides clues to the reason of the broadened transition.

The as-prepared sample was characterized by a routine resistance vs. temperature (R(T)) measurement and first stored in a desiccator, i.e. a sealed jar containing anhydrous calcium sulfate. No measurable change in the R(T) behavior was observed during two weeks of initial storage. The sample was then successively submerged in filtered, de-ionized water (ρ ≥ 18.2 MΩcm) at 22 °C water first for 1 hour, then for 3 hours, 6 hours, and 5 hours – thus at the end of each experiment, the sample had been exposed for a total of 1 hour, 4 hours, 10 hours, and 15 hours. After each water exposure, the sample was immediately dried with a jet of nitrogen gas and placed again in the desiccator for at least 10 minutes before characterization. Following this drying procedure, a R(T) curve was again recorded from room temperature to 11 K, and a small portion of the sample was cleaved off and analyzed by scanning electron microscopy (SEM) within 2 hours. With this procedure, all the data presented in this work is obtained on one single sample, eliminating possible errors resulting from sample-to-sample variations.

### 3. Results

The R(T) curves (normalized for clarity at 40 K) of the as-prepared sample and the same sample after water-exposure of various time intervals are shown in Figure 1. Transition temperatures and resistance values are summarized in Table 1. The as-prepared sample was characterized by a room-temperature resistance of 1Ω, corresponding to a resistivity of 80μΩcm. As we will show below, water exposure results in an inhomogeneous layer of partially decomposed material, making it difficult to define an effective film thickness; hence, the data are discussed in terms of resistance rather than resistivity. The room-temperature resistance increases drastically with exposure time, growing by more than two orders of magnitude for the entire sequence. Despite this drastic and rapid change, it is most remarkable that all of these samples have an identical $T_c^{onset}$ ~ 39 K, with a clear indication of a partial transition even in the sample exposed for a total of 15 hours and not showing zero-resistance above 11 K.

**Table 1.** Resistance at 300 K and 40 K, $T_c^{onset}$ and $T_{c0}$ for as-prepared MgB$_2$ and the same sample exposed to de-ionized water for a total of 1 hour, 4 hours, 10 hours, and 15 hours.

| Time (hour) | R$_{300K}$ (Ω) | R$_{40K}$ (Ω) | $T_c^{onset}$ (K) | $T_{c0}$ (K) |
|---|---|---|---|---|
| 0 | 1.08 | 0.63 | 39 | 31 |
| 1 | 1.28 | 0.77 | 39 | 27.6 |
| 4 = 1 + 3 | 1.51 | 0.99 | 39 | 25 |
| 10 = 1 + 3 + 6 | 7.66 | 5.51 | 39 | 17.8 |
| 15 = 1 + 3 + 6 + 5 | 153.8 | 119.2 | 39 | - |

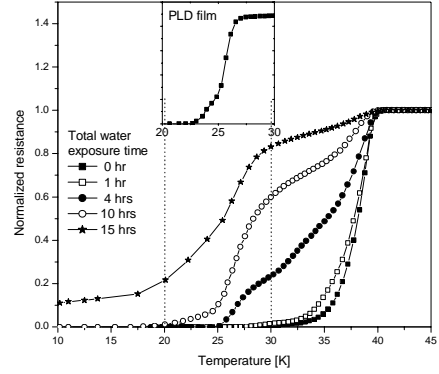

**Figure 1.** Resistance as function of temperature for as-prepared and water-exposed MgB$_2$ films on Al$_2$O$_3$. The inset shows data taken from Ref. [7] for a MgB$_2$ layer derived from a PLD-grown precursor. For clarity, the data are normalized at 40 K. All samples exhibit the same $T_c^{onset}$. A second transition at a temperature similar to $T_c$ for the PLD-grown film is clearly observed.

Inspecting the curves in Figure 1 more carefully, it is readily observed that there is a temperature at which the R(T) curves for the samples exposed to water for more than 1 hour show a negative curvature ($d^2R/dT^2 < 0$), indicative of a transition occurring at this lower temperature. For comparison, the inset of Figure 1 shows the data taken from Ref. 7. Those films were obtained by *in-situ* annealing of PLD-grown Mg-rich precursors, and have been shown to contain considerable amounts of oxygen [10].

Figure 2 shows SEM plan view and cross-section images of the sample as-prepared (a,b), after 4 hours of water exposure (c,d), and after 15 hours of water exposure (e,f). Figure 2a clearly shows well-defined, plate-shaped crystallites near the surface of the film, also visible in the cross-section image of Figure 2b (region labeled I). The same cross-section image further demonstrates that the sample exhibits some porosity in region II, as is usually observed in these materials except when annealed for a longer time [10]. There also appears to be a dense bottom layer (labeled III) in direct contact with the Al$_2$O$_3$ substrate. As we have shown previously [10], regions II and III of these films typically contain an amount of oxygen comparable to that observed in samples derived from PLD precursors, while the grains at the film surface are comparatively pure (no oxygen content within the detection limit of wave-length dispersive spectroscopy). The white spots in the image are interpreted as being elemental magnesium.

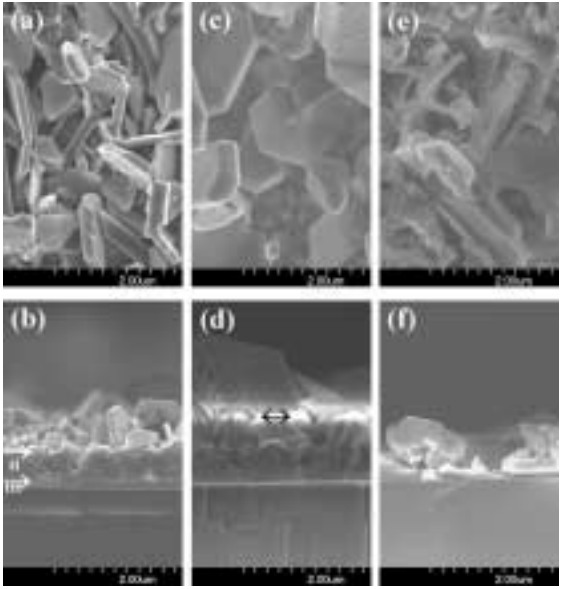

**Figure 2.** SEM images (plan-view and cross-section) of MgB$_2$ films. (a) and (b) show the as-prepared film, (c) and (d) the same sample after immersion into de-ionized water at room temperature for 4 hours, and (e) and (f) after immersion for 15 hours. Three distinctly different regions are labeled in Figure 2b by I, II, and III, as described in the text. The double arrow in Figure 2d indicates a region where strong charging was observed during exposure to the electron beam in the SEM, indicative of an insulating region.

As is clearly seen in Figure 2c, a 4-hour exposure to water results in the removal of the smaller crystallites at the film surface, with the larger platelets remaining but now exhibiting rounded edges. Acquiring a cross-section image was complicated due to some charging near the film surface (indicated by a double arrow), indicating that the resistance of portions of the material has increased drastically. Furthermore, the white spots clearly seen in the as-prepared sample and interpreted as elemental Mg are no longer present, while the dense layer closest to the interface remains unchanged.

Finally, the surface of the sample after exposure to water for 15 hours appears strongly corroded, the entire cross-section image exhibits charging effects, and the material starts to peel off from the substrate.

## 4. Discussions

A simple interpretation in terms of two layers exhibiting different transition temperatures is fully consistent with the results from the SEM study and the data in Figure 1. It is quite obvious that even after 15 hours of exposure to water, a fraction of the sample's volume still undergoes a transition at 39 K. Most likely, the interior of the largest grains is responsible for this behavior. These large grains, for which no oxygen contamination was detected in EDS measurements [10], appear to be well-connected in the as-prepared sample. It is possible that excess elemental magnesium also acts as a link between these grains, and current path linking these grains will show insignificant resistance below the 39 K.

As the sample is exposed to water, the smaller of the bulk-like grains at the film surface start to disappear, and the current must flow through a significant portion of the layer beneath these well-defined crystallites. That lower-lying layer of the film, for which an oxygen contamination similar to that of PLD-derived films was detected [10], appears to exhibit a significantly lower $T_c$.

Finally, prolonged exposure of the material to water ultimately results in the destruction of a sufficient amount of superconducting material so that a continuous superconducting path no longer exists while some of the grains, apparently, still undergo a superconducting transition at 39 K. A chemical analysis of the resulting material has not yet been performed, but it can be assumed that it contains significant amounts of MgO and B$_2$O$_3$ (note that MgB$_2$ reacts with air to form B$_2$O$_3$ [19]).

## 5. Conclusions

In conclusion, the described experiments consisting of immersing a MgB$_2$ film into water for up to 15 hours shed light not only onto the issue of chemical stability of this new superconductor, but also illustrate the non-uniformities observed in these layers obtained by an *ex-situ* reaction of a B film with Mg vapor. The presented data are consistent with an interpretation in which only a fraction of the film undergoes a transition at $T_c$ = 39 K, with some of the material resembles the PLD-derived films with a lower $T_c$.

**Acknowledgments**
This research is sponsored by the U.S. Department of Energy under contract DE-AC05-00OR22735 with the Oak Ridge National Laboratory, managed by UT-Battelle, LLC, and by the DOE Office of Energy Efficiency and Renewable Energy, Office of Power Technologies – Superconductivity Program.

References
[1] Akimitsu J 2001 *Symp. on Transition Metal Oxides, 10 January 2001 (Sendai)*
[2] Nagamatsu J, Nakagawa N, Muranaka T, Zenitani Y and Akimitsu J 2001 *Nature* **410** 63
[3] Larbalestier D C, Cooley L D, Rikel M O, Polyanskii A A, Jiang J, Patnaik S, Cai X Y, Feldmann D M, Gurevich A, Squitieri A A, Naus M T, Eom C B, Hellstrom E E, Cava R J, Regan K A, Rogado N, Hayward M A, He T, Slusky J S, Khalifah P, Inumaru K and Haas M 2001 *Nature* **410** 186
[4] Kambara M, Hari Babu M, Sadki E S, Cooper J R, Minami H, Cardwell D A, Campbell A M and Inoue I H 2001 *Supercond. Sci. Technol.* **14** L5
[5] Hirsch J E 2001 *Preprint* condmat/0102115
[6] Canfield P C, Finnemore D K, Bud'ko S L, Ostenson J E, Lapertot G, Cunningham C E and Petrovic C 2001 *Phys. Rev. Lett.* **86** 2423
[7] Christen H M, Zhai H Y, Cantoni C, Paranthaman M, Sales B C, Rouleau C, Norton D P, Christen D K and Lowndes D H 2001 *Preprint* cond-mat/0103478, Accepted for publication in *Physica C*.
[8] Zhai H Y, Christen H M, Paranthaman M, Cantoni C, Sales B C, Rouleau C, Christen D K and Lowndes D H 2001 *American Physics Society, March Meeting,* Web address: http://www.aps.org/meet/MAR01/mgb2/talks3.html#talk54,12 March 2001 (Seattle)
[9] Paranthaman M, Cantoni C, Zhai H Y, Christen H M, Aytug T, Sathyamurthy S, Specht E D, Thompson J R, Lowndes D H, Kerchner H R and Christen D K 2001 *Preprint* cond-mat/0103569, accepted for publication in *Appl. Phys. Lett.*
[10] Zhai H Y, Christen H M, Zhang L, Paranthaman M, Cantoni C, Sales B C, Christen D K and Lowndes D H 2001 *Preprint* cond-mat/0103618
[11] Zhai H Y, Christen H M, Zhang L, Cantoni C, Paranthaman M, Sales B C, Christen D K and Lowndes D H 2001 *Preprint* cond-mat/0103588
[12] Eom C B, Lee M K, Choi J H, Belenky L, Song X, Cooley L D, Naus M T, Patnaik S, Jiang J, Rikel M, Polyanskii A, Gurevich A, Cai X Y, Bu S D, Babcock S E, Hellstrom E E, Larbalestier D C, Rogado N, Regan K A, Hayward M A, He T, Slusky J S, Inumaru K, Haas M K and Cava R J 2001 *Preprint* cond-mat/0103425
[13] Kang W N, Kim H-J, Choi E-M, Jung C U and Lee S-I 2001 *Preprint* cond-mat/0103179
[14] Blank D H A, Hilgenkamp H, Brinkman A, Mijatovic D, Rijnders G and Rogalla H 2001 *Preprint* cond-mat/0103543
[15] Yan M F, Barns R L, O'Bryan Jr. H M, Gallagher P K, Sherwood R C and Jin S 1987 *Appl. Phys. Lett.* **51** 532
[16] Aytug T, Kang B W, Yan S L, Xie Y Y and Wu J Z 1998 *Physica C* **307** 117
[17] Jin S G, Zhu Z Z, Liu L M and Huang M L 1990 *Solid State Commun.* **74** 1087
[18] Zhou J P and McDevitt T 1992 *Chem. Mater.* **4** 953
[19] Callcott T A, Lin L, Woods T T, Zhang G P, Thompson J R, Paranthaman M and Ederer D L 2001 *Preprint* cond-mat/0103593

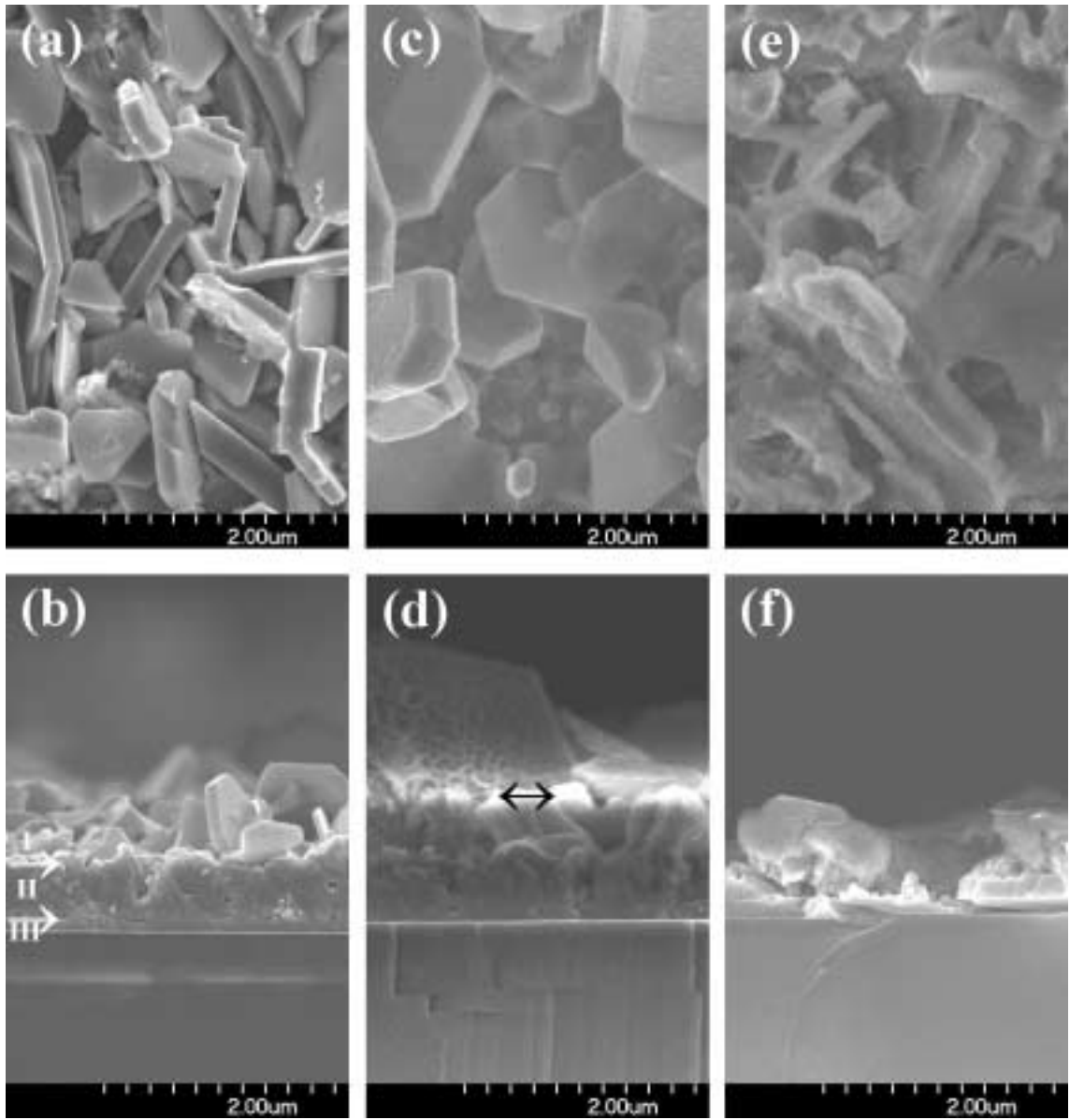

**Figure 2.** SEM images (plan-view and cross-section) of MgB$_2$ films. (a) and (b) show the as-prepared film, (c) and (d) the same sample after immersion into de-ionized water at room temperature for 4 hours, and (e) and (f) after immersion for 15 hours. Three distinctly different regions are labeled in Figure 2b by I, II, and III, as described in the text. The double arrow in Figure 2d indicates a region where strong charging was observed during exposure to the electron beam in the SEM, indicative of an insulating region.